\documentstyle[aps,preprint,prc,12pt]{revtex}
\begin{document}

\title{
Signals of a Critical Behavior in Peripheral Au + Au Collisions 
at 35 MeV/nucleon
}

\author{
{\underline{P. F. Mastinu$^{1,2}$}}, M. Belkacem$^{1,3}$, D. R. Bowman$^4$, 
M. Bruno$^1$, 
 M. D'Agostino$^1$,
J. D. Dinius$^5$,
A. Ferrero$^{6,7}$, M. L. Fiandri$^1$, C. K. Gelbke$^5$, T. Glasmacher$^5$,
F. Gramegna$^8$, D. O. Handzy$^5$, D. Horn$^4$, W. C. Hsi$^5$, M. Huang$^5$,
I. Iori$^6$, G. J. Kunde$^5$,
M. A. Lisa$^5$, W. G. Lynch$^5$,
G. V. Margagliotti$^{9}$, P. M. Milazzo$^{9}$, 
C. P. Montoya$^5$, A. Moroni$^6$, G. F. Peaslee$^5$,
F. Petruzzelli$^6$, R. Rui$^{9}$,
C. Schwarz$^5$, M. B. Tsang$^5$, G. Vannini$^{9}$, C. Williams$^5$,
V. Latora$^3$ and A. Bonasera$^3$
}

\address{
$^{1}$ Dipartimento di Fisica and INFN, Bologna, Italy \\
$^{2}$ Dipartimento di Fisica, Padova, Italy \\
$^{3}$ INFN, laboratorio Nazionale del Sud, Catania, Italy \\
$^{4}$ Chalk River Laboratories, Chalk River, Canada \\
$^{5}$ NSCL, Michigan State University, USA \\
$^{6}$ Dipartimento di Fisica and INFN, Milano, Italy \\
$^{7}$ CNEA, Buenos Aires, Argentina \\
$^{8}$ INFN, Laboratori Nazionali di Legnaro, Italy \\
$^{9}$ Dipartimento di Fisica and INFN, Trieste, Italy
}

\maketitle


\vspace{0.2 cm}
\centerline{ (MULTICS - MINIBALL Collaboration) }
\vspace{0.2 cm}


\begin{abstract}
\baselineskip=14pt

Multifragment events resulting from peripheral Au + Au collisions at 35
MeV/nucleon 
are analysed in terms of critical behavior. The analysis of most of
criticality signals 
proposed so far (conditional moments of charge distributions, Campi scatter
plot, 
fluctuations of the size of the largest fragment, intermittency analysis) is
consistent 
with the occurrence of a critical behavior of the system.

\end{abstract}


\newpage
\baselineskip=15pt


Initiated
by the observation of fragments in the final stages of the reaction
exhibiting a power law in
fragment
charge distributions \cite{eos1}, and stimulated by the similarity of
nuclear matter equation of state with that of a van der Waals gas \cite{nm_eos},
the possibility of observing a liquid-gas phase
transition in nuclear systems has been the subject of intensive investigations
(experimental and theoretical) for more than a decade
\cite{lg1,lg2,belkacem}.
This interest increased recently with the attempt by the EOS Collaboration to
extract the critical exponents of fragmenting nuclear systems produced in the
collision of 1 GeV/nucleon Au nuclei with a carbon target \cite{eos2}, and
with the extraction by the ALADIN Collaboration of the caloric curve resulting
from the fragmentation of the quasiprojectile formed in the collision
Au + Au at 600 MeV/nucleon which exhibits a behavior suggested for a liquid-gas
phase transition \cite{aladin}.

In this contribution, we report the results of a recent
experiment conducted by the
MULTICS-MINIBALL Collaboration in which we studied the fragmentation
(determined on an event-by-event basis) resulting from peripheral
Au + Au collisions at E = 35 MeV/nucleon.
The study of this reaction within the framework of Classical Molecular
Dynamics (CMD) 
model indicates the possible occurrence of a critical behavior in 
peripheral collisions
as reported in Ref.
\cite{belkacem1,theo}. Following this investigation the experimental data are
analyzed in terms of critical behavior.

The experiment has been performed at the National Superconducting Cyclotron 
Laboratory of the Michigan State University using the coupled Multics-Miniball
apparatus which has a geometric acceptance greater than 87\% of 4$\pi$. The
Multics 
apparatus covered angles between 3 and 23 degrees in the laboratory frame and 
detected fragments with charge up to $Z=83$, with an energy threshold of about
1.5 
MeV/nucleon, independently of fragment charge \cite{strum}. Charged particles
with 
charge up to $Z=20$ were detected at $23 \leq \theta _{lab} \leq 160$ by 159 
phoswich detector elements of the MSU Miniball \cite{mini} with energy 
thresholds of about 2, 3, and 4 MeV/nucleon for $Z=3, 10, 18$, respectively.

Guided by the previously cited CMD calculations, semi-peripheral 
and peripheral events are identified selecting the events where the largest 
fragment has the velocity component along the beam direction 
greater than 75\% of the beam velocity and the total
detected 
charge between 70 and 90.

Figure 1 shows a scatter plot of $ln(Z^{j}_{max})$ versus $ln(M^{j}_{2})$
("Campi scatter plot" \cite{campi1}) where $Z^{j}_{max}$
is the charge of the heaviest fragment and $M^{j}_{2}$
is the second conditional moment of the charge distribution
detected in the $j$-th event,
\begin{equation}
M^{(j)}_{2} = \sum_{Z} Z^{2} n_{j}(Z)
\label{mom}
\end{equation}
Here, $n_{j}(Z)$ denotes the number of fragments of charge $Z$
detected in the $j$-th event, and the summation is over all
fragments but the heaviest detected one. Theoretical
investigations suggest that such plots may be useful in
characterizing near-critical behavior of finite systems \cite{campi1}.
The calculated Campi scatter plots typically exhibit two
branches: an upper branch with a negative slope
containing largely undercritical events (e.g. $T < T_{crit}$ in a
liquid-gas phase transition or $p > p_{crit}$ in a percolation
phase transition) and a lower branch with a positive slope
containing largely overcritical events ($T > T_{crit}$ or $p <
p_{crit}$). The two branches were shown to meet close to the
critical point of the phase transition \cite{belkacem,campi1,gross1}.

The data shown in Fig. 1, display two branches similar to
the ones predicted for undercritical and overcritical
events. In the top-right part, close to the intersection of
these two branches (potentially containing near-critical
events), a separate island is observed which is due to
fission events, as first noted by Ref. \cite{gross1}. By appropriate
gates in the Campi plot, these fission events are removed
from the following analysis.

To further investigate the two branches observed in Fig. 1
and the region where they intersect, we employ three cuts
selecting the upper branch (cut 1), the lower branch (cut
3) and the intersection region (cut 2); these cuts are
indicated in Fig. 1. The charged particle multiplicity
distributions observed for these three cuts are shown as
dashed histograms in the upper part of Fig. 2 
together with the multiplicity distribution
obtained for the totality of the selected events (solid histogram). 
Cuts 1 and 3 largely select
low and high multiplicity events corresponding to very
peripheral and more central collisions (assuming on the average
a monotonic
relation between $N_{c}$ and impact parameter); cut 2
represents a wide range of charged particle multiplicities
and thus may involve a wide range of intermediate impact
parameters. Thus emission from a unique source cannot
be ascertained for cut 2, and it is likely that this cut
contains contributions from projectile and target-like
sources and from the neck \cite{montoya} which emits lighter
fragments with enhanced probability as
compared to the projectile residue \cite{montoya}.
However, one cannot exclude that this large multiplicity distribution 
be related to the occurrence of large fluctuations as expected at the critical
point. Moreover reducing the size of the region 2 does not change the large
multiplicity distribution. 
In the lower part of Fig. 2 the multiplicity distribution given by cut 2
is reported as a solid line, while the dashed lines represents 
the multiplicity distribution obtained splitting this cut 2 
in two equal parts in the vertical direction. No substantial changement is 
noticed.

Fragment charge distributions, not corrected for detection
efficiency, are presented \cite{note} for the three cuts in the upper panels
of Fig. 3 (cut 1: left panel; cut 2: central panel and cut 3: right panel).
Cut 1 contains light fragments and heavy
residues and thus resembles the "U"-shaped distributions
predicted by percolation calculations in the sub-critical
region \cite{bauer}. For cut 3, one observes an
unusually flat charge distribution similar to the one
previously reported \cite{central} for central collisions which were
selected without the specific constraints employed in this
paper and attributed to a Coulomb-driven breakup of a
very heavy composite system \cite{central} (The steep fall-off at
large $Z$ is an artifact of the selection of events with $Z = 70 - 90$
used in this paper). For cut 2, a
fragment charge distribution is observed which resembles
a power-law distribution, $P(Z) \propto Z^{-\tau}$, with $\tau \approx 2.2$.
For macroscopic systems exhibiting a liquid-gas phase
transition, such a power-law distribution is predicted to
occur near the critical point \cite{fisher}. However, it is not yet
known by how much the final fragment distributions
differ from the primary ones after the sequential decays
of particle unstable primary fragments.

The lower panels of Fig. 3 show \cite{note} the logarithm of the Scaled 
Factorial
Moments (SFM), defined as
\begin{equation}
F_i(\delta s)={{\sum _{k=1}^{Z_{tot}/ \delta s}<{n_k}\cdot ({n_k}-1)\cdot ...
\cdot({n_k}-i+1)>}
\over {\sum _{k=1}^{Z_{tot}/ \delta s}<n_k>^i}}
\label{SFM}
\end{equation}
($i$ = 2, ..., 5), as a function of the logarithm of the bin size
$\delta s$. In the above definition of the SFM, $Z_{tot} = 158$ and $i$ is
the order of the moment. The total interval $[1, Z_{tot}]$ is
divided into $M = Z_{tot}/\delta s$ bins of size $\delta s$, $n_k$ is the
number
of particles in the $k$-th bin for an event, and the brackets $< >$
denote the average over many events. A linear rise of
the logarithm of the SFM versus $\delta s$ (i.e.
$F_{i} \propto \delta s^{-\lambda_{i}}$) indicates an
intermittent pattern of fluctuations \cite{gross1,bialas,plocia}. Even
though this quantity is ill defined for fragment
distributions \cite{campi2,phair,delzoppo}, several theoretical studies have
indicated that critical events give a power law for the SFM
versus the bin size \cite{belkacem,plocia}. For cut 3 (right part of the
figure), the logarithm of the scaled factorial moments
$ln(F_{i})$ is always negative (i.e. the variances are smaller
than Poissonian \cite{plocia}) and almost independent on $\delta s$; there
is no intermittency signal. The situation is different for cut
2 (central part). The logarithm of the scaled factorial
moments are positive and almost linearly increasing as a
function of $-ln(\delta s)$, and an intermittency signal is observed.
Region 1, corresponding to evaporation, gives zero slope.
Increasing or reducing the size of the three cuts in the
respective regions does not change significantly these
results. The interpretation of experimentally observed
intermittency signals may, however, be problematic due
to ensemble averaging effects \cite{phair}, even though calculations show that
impact parameter averaging only increases the absolute value of the SFM
\cite{delzoppo}.
Since cut 2 may involve a
large range of impact parameters, the observed
intermittency signal could be an artifact of ensemble
averaging and can, therefore, not be taken as a definitive
proof of unusually large fluctuations in a sharply defined
class of events.

In the upper part of Fig. 4, we have
plotted the second moment $M_{2}$ versus the multiplicity of charged particles
$N_{c}$. In a macroscopic thermal system, $M_{2}$ is proportional to 
the isothermal compressibility which diverges at the critical temperature
\cite{balescu,finocchiaro}. Of
course, in finite systems, the moments remain finite due to finite size
effects. The moment $M_{2}(N_{c})$ is averaged over all the events having the
multiplicity $N_{c}$, except those corresponding to fission (see above).
One sees
that the second moment shows a "peak" for a multiplicity $N_{c}$ around 20-25.
The EOS Collaboration has found a "critical" multiplicity of 26, somewhat
higher than our result \cite{eos2}.
We stress, however, that, because for this experiment
the Multics apparatus was set not to well detect Z$<$3 charges, some particles
(protons and alphas) are lost. Thus our results are not in contrast with those
of EOS. 

Another quantity introduced by Campi to characterize the critical behavior of
a system is the relative variance $\gamma_{2}$ defined as \cite{campi2,plocia}:
\begin{equation}
\gamma_{2} = \frac{M_{2}M_{0}}{M_{1}^{2}}
\label{gam2}
\end{equation}
It was shown by Campi that this quantity presents a peak around the
critical point which means that the fluctuations in the fragment size
distributions are large near the critical point \cite{campi2}.
The lower part of Fig. 4 shows the plot of $\gamma_{2}$ versus charged particle
multiplicity $N_{c}$. This quantity shows a peak for 
$N_{c} \approx 20-25$, consistent with that observed for $M_{2}$.

The upper panel of Fig. 5 shows the second moment of charge
distribution $M_{2}$ 
calculated according to Eq. (\ref{mom}) by excluding the largest fragment
(solid line) and including it as suggested by
Stauffer and Aharony \cite{stauffer}. Physically, $Z_{max}$ corresponds to the
bulk in the liquid region and thus should be subtracted from the calculation
of the moments, but not in the gas region where the bulk no more exists. 
This plot must be compared with the lower panel of the figure which 
reports the results of simulations done by Bauer and
Botvina \cite{bauer1} for the reaction Au + C at 1 GeV/nucleon
based on INC+SMM model and on INC+Percolation models. This figure reports also 
the experimental data of the EOS 
collaboration (open circles). See Ref. \cite{bauer1} for details.
The shape of this quantity 
resembles that of our experimental data. One notes that our data show the 
trend very similar to that of the OES data and the simulations.
 
Another quantity recently proposed so far as a signal for criticality is 
the normalized variance of the charge of the largest fragment $\sigma_{NV}$. 
This quantity, defined by:
\begin{equation}
\sigma_{NV} = \frac{\sigma^{2}_{Z_{max}}}{<Z_{max}>}
\label{nv}
\end{equation}
where
\begin{equation}
\sigma^{2}_{Z_{max}} = <Z_{max}^{2}> - <Z_{max}>^{2}
\end{equation}
(the brackets $<~^.~>$ indicate an ensemble averaging)
shows a peak at the critical point, where charge 
distributions are expected to show the largest fluctuations.
Figure 6 shows in the upper part the size of the largest fragment, and in the 
lower part
the $\sigma_{NV}$ for our data versus charged particle 
multiplicity. This plot shows a clear peak for multiplicities around
$N_{c} = 20$ as expected around the critical point, indicating that charge
distributions show the largest of fluctuations at this multiplicity.

In conclusion, we have analyzed fragment production in
Au + Au collisions at E = 35 MeV/nucleon. Events were selected
by requiring a total detected charge between 70 and 90
and the velocity of the largest detected fragment larger
than 75\% of the projectile velocity. A Campi scatter plot of
these events displays two branches similar to the sub-
and overcritical branches observed in theoretical studies.
The selection of events from the intersection of these two
branches (which has been associated with critical events
in theoretical studies) shows a power law charge
distribution with an exponent of $\tau\approx 2.2$ similar to that
characterizing the mass distribution near the critical point
of a liquid-gas transition. These events, further, display
an intermittent behavior similar to that expected for
near-critical events. Moreover, the second moment of fragment
charge distribution $M_{2}$, the relative variance $\gamma_{2}$ and the 
normalized variance of the size of the largest fragment show peaks 
indicating large fluctuations for 
multiplicities around $N_{c} = 20$, as expected near the critical point. 
While these signatures have been
associated with near-critical events, we must caution that
the effects of finite experimental acceptance and event
mixing with possible contributions from the decay of
projectile-like fragments and the neck-region are not yet
sufficiently well understood to allow an unambiguous
conclusion of critical behavior in the present reaction. Our
work does, however, show that different regions of the
nuclear phase diagram can be probed at one incident
beam energy by selecting events according to different
impact parameters and/or energy depositions.




\begin{figure}
\label{f1}
\noindent
\caption{Campi scatter plot, $ln(Z_{max})$ versus $ln(M_{2})$.
The three different regions are discussed in the text. Fission events are 
to the right of region 2.} 
\end{figure}

\begin{figure}
\label{f2}
\noindent
\caption{Upper panel: Multiplicity distribution of the 
events selected for the analysis (solid histogram). The three dashed 
histograms (1, 2 and 3) represent the multiplicity 
distributions of the events falling in the three cuts drawn in Fig. 1.
Lower panel: Multiplicity distribution for cut 2 in Fig. 1 (solid histogram). 
Dashed histograms show $N_{c}$-distributions when this cut 2 is split 
vertically in two equal parts.}
\end{figure}
 
\begin{figure}
\label{f3}
\noindent
\caption{Charge distributions and the corresponding scaled factorial moments
$ln(F_{i})$ versus $- ln(\delta s)$ for the events falling in the three cuts in
the Campi scatter plot; the left part of the figure corresponds to cut 1, the
central part to cut 2 and the right part to cut 3. The line
in the upper central part represents the power law $Z^{-\tau}$
with $\tau=2.2$. In the lower part, solid points represent the SFM of order
$i=2$, open circles
$i=3$, open squares $i=4$, and open triangles $i=5$.} 
\end{figure}

\begin{figure}
\label{f4}
\noindent
\caption{Upper panel: Second moment $M_{2}$ of charge distribution versus
multiplicity $N_{c}$. Lower panel: Relative variance $\gamma_{2}$.}
\end{figure}

\begin{figure}
\label{f5}
\noindent
\caption{Upper panel: Second moment $M_{2}$ of charge distribution versus
multiplicity distribution calculated omitting the biggest fragment (lower curve)
and taking it into account (upper curve). Lower panel: Same as upper panel. 
This figure is taken from Ref. [27]. Open circles represent 
experimental data of the EOS Collaboration [6]. For the description of 
solid lines and histograms in the lower panel see Ref. [27].}
\end{figure}

\begin{figure}
\label{f6}
\noindent
\caption{Size of the largest fragment (upper panel), and
Normalized variance of the charge of the biggest fragment 
$\sigma_{NV}$ (lower panel) versus multiplicity of charge particles $N_{c}$.}
\end{figure}


\begin{thebibliography}{99}
\bibitem{note} The charge distributions shown in Fig. 3 and 
the SFM shown in Fig. 4 have been calculated without the heaviest fragment
in each event, apart the charge distribution of cut 1.
\end{thebibliography}
\end{document}